\documentclass[a4paper,11pt]{amsart}

\usepackage{amsmath}
\usepackage{amssymb}
\usepackage{amsthm}
\usepackage{amscd}
\usepackage{amsaddr}
\usepackage{enumitem}   
\usepackage{hyperref}

\newcommand{\g}{\mathfrak{g}}
\newcommand{\h}{\mathfrak{h}}

\title[Godbillon-Vey as a Casimir in Ideal Fluids]{The Godbillon-Vey Invariant as a Restricted Casimir of Three-dimensional Ideal Fluids}
\author{Thomas Machon}
\address{H.H.~Wills Physics Laboratory, Tyndall Avenue, Bristol BS8 1TL, UK}
\begin{document}

\maketitle

\begin{abstract}
We show the Godbillon-Vey invariant arises as a `restricted Casimir' invariant for three-dimensional ideal fluids associated to a foliation. We compare to a finite-dimensional system, the rattleback, where analogous phenomena occur.
 \end{abstract}

\section{Introduction}

The topological aspect of ideal fluids has its origins in the transport of vorticity. A consequence is the conservation of helicity, which measures the average linking of vortex lines~\cite{moffatt69, arnold74, moffatt92}. In the Hamiltonian formulation of ideal fluids as an infinite-dimensional Lie-Poisson system, helicity appears as a Casimir invariant, a degeneracy in the Lie-Poisson bracket~\cite{morrison98}.

The state of an ideal fluid on a homology 3-sphere is specified by the vorticity, a divergence-free vector field. A Casimir in an ideal fluid is invariant under all volume-preserving diffeomorphisms of the domain, so can be said to measure a topological property of the vorticity. For a generic vorticity field, helicity is the only topological invariant~\cite{enciso16, kudryavtseva14, kudryavtseva16}. However, higher order invariants can be defined in special cases. Here we study the Godbillon-Vey invariant, $GV$, which can be associated to a vorticity field tangent to a codimension-1 foliation~\cite{machon20, webb14, webb19, tur93}. $GV$ originates in the theory of foliations~\cite{gv,candel}; in ideal fluids it measures topological helical compression of vortex lines~\cite{machon20}. The goal of this paper is to show how $GV$ fits naturally into the Lie-Poisson Hamiltonian formulation of ideal fluids~\cite{morrison98} as a `restricted Casimir' invariant. In particular, we consider a set $S$ of ideal fluids where the Lie-Poisson bracket has an additional degeneracy associated to the Lie subalgebra of volume-preserving vector fields tangent to a foliation, which may vary within $S$. On $S$ we construct a modified Lie-Poisson type bracket, in terms of which the Godbillon-Vey invariant appears as a Casimir.

Imposing this degeneracy also forces the helicity to vanish and in this sense $GV$ is hierarchical, in a manner analogous to that suggested by Arnold and Khesin~\cite{arnold99}. Recent work~\cite{yoshida14,yoshida16} has studied similar hierarchical structures in Hamiltonian systems, where a singular region in phase space with a Poisson operator of decreased rank can itself be considered as a Poisson submanifold, on which new Casimir invariants appear. What we describe can be considered an example of this phenomenon. We note also that a foliation on $M$ can be defined through an exact sequence of vector bundles
\begin{equation}
\begin{CD}
0@>>> T\mathcal{F} @>>> TM @>>>N @>>>0
\end{CD}
\end{equation}
where $N = T\mathcal{F}/TM$ is the normal bundle of the foliation. This sequence passes to the volume-preserving case, and we note the connection to the classification of Casimir invariants coming from Lie algebra extensions~\cite{thiffeault00}. 

There is a finite-dimensional example in the Lie-Poisson formulation of the `rattleback' spinning top~\cite{yoshida17}, where corresponding phenomena occur: there is a submanifold of phase space where the Poisson operator has an additional degeneracy associated to a Lie subalgebra; on this submanifold the primary Casimir vanishes and a new restricted Casimir appears.

In the finite-dimensional rattleback case, perturbation of the system around the singular manifold leads to interesting dynamical properties~\cite{yoshida17}. Our own analysis of the Godbillon-Vey invariant elsewhere also suggests a strong connection to dynamics; $GV$ provides a global and local obstruction to steady flow and can be used to estimate the rate of change of vorticity~\cite{machon20}. With that in mind, we suggest that flows with $GV\neq 0$ (or perturbations thereof) may prove particularly interesting from a dynamical perspective. Finally, we note our light touch regarding rigour.

\section{Lie-Poisson Systems} 

See e.g.~\cite{thiffeault00} for a description. Let $\g$ be a Lie algebra associated to a group $G$, with $\g^\ast$ its dual. Given an element $\alpha \in \g^\ast$ and two elements $U,V \in \g$ we form the bracket
\begin{equation}
\langle \alpha , [U,V] \rangle,
\label{eq:KKS}
\end{equation} 
where $\langle \cdot, \cdot \rangle : \g^\ast \times \g \to \mathbb{R}$ is the natural pairing between the Lie algebra and its dual, and $[ \cdot, \cdot]$ is the Lie bracket of $\g$. This is then used to define the Lie-Poisson bracket
\begin{equation}
\{F,G\}_\pm = \pm \left \langle \alpha , \left [ \frac{\delta F}{\delta \alpha}, \frac{\delta G}{\delta \alpha} \right ] \right \rangle,
\label{eq:pb}
\end{equation} 
where the (functional) derivative $\delta F/ \delta \alpha$ is identified with an element of $\g$ by the relation
\begin{equation}
\frac{d}{d \epsilon} F(\alpha + \epsilon \delta \alpha) \big |_{\epsilon=0} =\left  \langle \delta \alpha , \frac{\delta F}{\delta \alpha} \right \rangle.
\end{equation}
The sign in \eqref{eq:pb} depends on whether we consider right-invariant or left-invariant function(al)s on $\g^\ast$ with respect to the coadjoint representation of $G$, but is irrelevant for our purposes. Coupled with a Hamiltonian function on $\g^\ast$, this specifies the system. The noncanonical nature of the Lie-Poisson bracket allows for the existence of Casimir invariants, $C$, given by the property $\{F,C\}=0$ for any function $F$. We define the coadjoint bracket $[ \cdot, \cdot]^\dagger : \g \times \g^\ast \to \g^\ast$ as
\begin{equation}
\langle [U, \alpha]^\dagger , V \rangle = \langle \alpha, [U,V] \rangle.
\end{equation}
This allows us to give the condition for $C$ to be a Casimir as \begin{equation}
\left [  \frac{\delta C(\alpha)}{\delta \alpha} , \alpha \right ]^\dagger =0.
\end{equation}

In this paper we will be interested in sets of points $\alpha \in S \subset \g^\ast$ where there is a non-generic degeneracy associated to a subalgebra $\mathfrak{h}_\alpha \subset \g$, such that 
\begin{equation}
\langle \alpha, U \rangle =0,
\end{equation}
for $U \in \mathfrak{h}_\alpha$. For a given $\alpha \in \g^\ast$, let $\beta = \textrm{ad}^\ast_g \alpha$, $g \in G$. Then $\beta$ is orthogonal to the subalgebra $\h_\beta = \textrm{ad}_g \h_\alpha $, so that $S$ will, in general, be a set of coadjoint orbits in $\g^\ast$. The precise specification of admissible subalgebras and the subset $S$ in a general formulation is left intentionally vague. 

\section{Finite Dimensional Example: the Rattleback}

An idealised description of the chiral dynamics of the rattleback spinning top~\cite{yoshida17} can be formulated as a Lie-Poisson system based on the three-dimensional Lie algebra with Bianchi classification $\textrm{VI}_{h<-1}$, spanned by three elements, $P$, $R$, $S$ with Lie bracket
\begin{equation}
[P,R] =0, \quad [S,P] = h P, \quad [S,R] = R.
\end{equation}
Physically $P$, $R$, and $S$ are associated to pitching, rolling and spinning motions respectively, and $h$ is a geometric parameter related to the aspect ratio of the top. The dynamical variable is an element of the dual space $\g^\ast$ which we write as a lowercase tuple $(p,r,s)$, in terms of which the dynamics are~\cite{yoshida17, moffatt08}
\begin{equation}
\frac{d}{dt} \begin{pmatrix} p \\ r \\ s \end{pmatrix} = \begin{pmatrix} -h ps \\ - rs \\ r^2 +h 
p^2\end{pmatrix}.
\label{eq:rattleback}
\end{equation}
The Hamiltonian of this system is given by $H = (p^2+r^2+s^2)/2$. At a generic point in $\g^\ast$ the Lie-Poisson bracket has a one-dimensional kernel, associated to the Casimir
\begin{equation}
C = pr^{-h},
\end{equation}
which one can check is conserved by the dynamics \eqref{eq:rattleback}. 

There is a two-dimensional Abelian subalgebra $\h \subset \textrm{VI}_{h<-1}$, spanned by $P,R$. The set of points $ M \subset \g^\ast$ orthogonal to $\h$ is the singular line $(0,0,s)$, so that on $M$ the Casimir $C=0$. On $M$ the Lie-Poisson bracket is trivial, so the dynamics are trivial (one can see this by setting $p=r=0$ in \eqref{eq:rattleback}). It follows that $s$ is a constant of the motion on $M$ only. Finally, note that $M$ can be thought of as a one-dimensional Poisson manifold with trivial Poisson bracket, and with respect to this bracket $s$ is a Casimir invariant (as is any function of $s$), so that $s$ is a restricted Casimir invariant of the rattleback system. Physically it corresponds to simple spinning motion of the top.

\section{The Godbillon-Vey Invariant as a Restricted Casimir in Three-Dimensional Ideal Fluids}

Now we see how the same pattern of phenomena is found in three-dimensional ideal fluids on a manifold $M$. We assume throughout that $M$ is a homology 3-sphere (one can take $M=S^3$).

\subsection{Ideal Hydrodynamics and Helicity}
\label{sec:hyd}

In the Lie-Poisson formulation of ideal fluids~\cite{morrison98,arnold99}, $\g$ is the Lie algebra of volume-preserving vector fields on $M$ with respect to a volume form $\mu$, so that $\mathcal{L}_U \mu =0$ for $U \in \g$ and the $2$-form $\iota_U \mu$ is closed. The dynamical variable is given by an element of the dual space $\g^\ast$, the smooth part of which can be identified as $\Omega^1(M)/d\Omega^0(M)$, the space of differential 1-forms modulo exact forms, and each element is given by a coset $[\alpha]$, with specific representative $\alpha$. We will suppress the coset notation $[]$. The pairing $\langle \cdot, \cdot \rangle: \g^\ast \times \g \to \mathbb{R}$ is given by 
\begin{equation}
\langle \alpha, U \rangle = \int_M (\iota_U \alpha) \mu,
\label{eq:pair}
\end{equation}
which does not depend on the representative 1-form $\alpha$. In this case the Lie-Poisson bracket takes the form
\begin{equation}
\langle \alpha, [U,V] \rangle = \int_M \alpha \wedge \iota_{[U,V]} \mu.
\label{eq:pair}
\end{equation}
The coadjoint bracket is given as
\begin{equation}
[U, \alpha]^\dagger = -\iota_U d \alpha = -\iota_U \iota_W \mu.
\end{equation}
Where the vorticity field $W \in \g$ is given by $d \alpha= \iota_W \mu$. Helicity is defined as
\begin{equation}
\mathcal{H} = \int_M \alpha \wedge d \alpha = \langle  \alpha, W \rangle.
\end{equation}
A short calculation gives
\begin{equation}
\frac{d}{d \epsilon} \mathcal{H}(\alpha + \epsilon \delta \alpha) \big |_{\epsilon=0} = \int_M \delta \alpha \wedge 2 d \alpha,
\end{equation}
so $\delta \mathcal{H}/ \delta \alpha = 2 W$ and hence
\begin{equation}
\left [  \frac{\delta \mathcal{H}}{\delta \alpha} , \alpha \right ]^\dagger = 0,
\end{equation}
so that $\mathcal{H}$ is a Casimir.

\subsection{Foliations and $\g^\ast$}

We now consider a codimension-1 foliation $\mathcal{F}_\alpha$ of $M$ such that $\alpha \in \g^\ast$ satisfies
\begin{equation}
\langle \alpha , X \rangle =0,
\end{equation} 
for $X$ in the subalgebra of volume-preserving vector fields $ \h_\alpha \subset \g$ that are tangent to the leaves of $\mathcal{F}_\alpha$. Let $\beta_\alpha$ be a defining form for $\mathcal{F}_\alpha$ and consider the family of closed 2-forms $d(h \beta_\alpha)$, $h$ a function, then the vector field $Y$ defined by $\iota_Y \mu = d(h \beta_\alpha)$ is an element of $\mathfrak{h}_\alpha$.  By assumption on $\alpha$,
\begin{equation}
0 = \int_M \alpha \wedge d (h \beta_\alpha) = \int_M h \beta_\alpha \wedge d \alpha.
\end{equation} 
As $h$ is arbitrary, $\beta_\alpha \wedge d \alpha =0$. Recall the vorticity field $W$ defined by $d \alpha = \iota_W \mu$, it follows that
\begin{equation}
W \in \mathfrak{h}_\alpha. 
\end{equation}
As an immediate consequence, the helicity, $\langle \alpha, W \rangle$, vanishes.
Now let $\gamma$ be a closed loop tangent to $\mathcal{F}$. The quantity
\begin{equation}
I_\gamma = \int_\gamma \alpha
\end{equation}
is invariant under leafwise homotopies of $\gamma$, so that $\alpha$ defines a class $[\alpha]_\mathcal{F}$ in the foliated cohomology group $H^1(\mathcal{F}_\alpha)$. In fact
\begin{equation}
[\alpha]_\mathcal{F} = 0 \in H^1(\mathcal{F}_\alpha).
\end{equation}
Consider a family of smooth vector fields $G_\lambda \in \g$ with support in a tubular neighbourhood of $\gamma$ of diameter $\sim \lambda$ (with respect to a metric), tending to the singular vector field with support $\gamma$ and constant flux $\phi$ as $\lambda \to 0$, so that $\int_D \iota_{G_\lambda} \mu \to \phi$ as $\lambda \to 0$, where $D$ is a disk pierced by $\gamma$. Then $\langle \alpha, G_\lambda \rangle \to \phi I_g$ as $\lambda \to 0$. But $|\langle \alpha, G_\lambda \rangle| < \lambda C$ for some constant $C$, so $I_g=0$. As $\gamma$ was arbitary, $[\alpha]_\mathcal{F} = 0 \in H^1(\mathcal{F}_\alpha)$. As an immediate consequence, any representative of the coset $[\alpha] \in \Omega^1(M) / d \Omega^0(M)$ can be written as $ f \beta_\alpha + dg$ for functions $f,g$. There is then a canonical representative form $\alpha_c = f \beta_\alpha$, which we write as $\alpha$ in subsequent sections. With this choice of representative the helicity density vanishes, $\alpha_c \wedge \iota_W \mu =0$.

\subsection{$\Omega^1_I(M)$ and the Functional Derivative}

We may write $\alpha = f \beta_\alpha$. For simplicity we will assume that $f \neq 0$, and we choose $\beta_\alpha$ so that $\alpha = \beta_\alpha$ is a defining form for $\mathcal{F}_\alpha$. Then, with this canonical choice, the subset of $\g^\ast$ we are considering can be identified with the space of nonvanishing integrable 1-forms on $M$, which we write as $\Omega^1_I(M)$ (this space is not connected, we consider a single arbitrary connected component). We note that $\Omega^1_I(M)$ is no longer a vector space.

We would like to define functional derivatives on $\Omega^1_I(M)$. For a one-parameter family $\alpha_t \in \Omega^1_I(M)$, $t \geq 0$ with time derivative $\dot{\alpha}_t$, we write $\dot{\alpha} = \dot{\alpha}_0$ and $\alpha = \alpha_0$. Now for a functional $F$ on $\Omega^1_I(M)$ the functional derivative is defined as
\begin{equation}
\frac{d}{d t} F(\alpha_t ) \big |_{t=0} =\left  \langle  \dot{\alpha} , \frac{\delta F}{\delta \alpha} \right \rangle,
\end{equation}
and we identify ${\delta F}/{\delta \alpha}$ with a vector field as in Section \ref{sec:hyd}. Because we have a canonical choice of $\alpha$, we no longer require invariance under gauge transformations and so are not restricted to volume-preserving vector fields. We suppose instead ${\delta F}/{\delta \alpha} \in \mathfrak{X}(M) / \Xi_\alpha $ where $\mathfrak{X}(M)$ is the space of smooth vector fields on $M$ and $\Xi_\alpha \subset \mathfrak{X}(M)$ is an $\alpha$ dependent subset satisfying $\langle \dot{\alpha}, U \rangle =0$ for $U \in \Xi_\alpha$. 

Our characterisation of $\Xi_\alpha$ below is not complete, but is sufficient for our purposes. First, we will show that it is non-empty. As $\alpha_t$ is integrable we have
\begin{equation}
\alpha_t \wedge d \alpha_t =0.
\end{equation}
In particular this gives
\begin{equation}
0 = \frac{d}{d t} \left( \int_M f \alpha_t \wedge d \alpha_t  \right )\Big |_{t=0}= \int_M \dot{\alpha} \wedge (f d \alpha + d (f \alpha))
\label{eq:degeneracy}
\end{equation}
for any function $f$, so that fields $V$ satisfying 
\begin{equation}
\iota_V \mu = f d \alpha + d (f \alpha)
\label{eq:what}
\end{equation}
are elements of $\Xi_\alpha$. Now we give two properties of general elements of $\Xi_\alpha$. 

Firstly we note that any field in $\Xi_\alpha$ must be tangent to $\mathcal{F}_\alpha$. We can choose $\alpha_t = \exp(g t) \alpha$, so that $\dot{\alpha} = g \alpha$ for an arbitrary function $g$. Now suppose $U$ is not tangent to $\mathcal{F}_\alpha$, then by an appropriate choice of $g$ we can force $\langle g \alpha, U \rangle \neq 0$, so $U \notin \Xi_\alpha$. 

Secondly we note that any element $V$ of $\Xi_\alpha$ must satisfy $d (\iota_V \mu)  = \eta \wedge (\iota_V \mu)$, where $\eta$ is a 1-form defined by the relation $d \alpha = \alpha \wedge \eta$. We can choose $
\alpha_t$ to be generated by a family of diffeomorphisms, so that $\dot{\alpha}= \mathcal{L}_U \alpha$ for $U \in \mathfrak{X}(M)$. Then, writing $\nu = \alpha \wedge \sigma = \iota_V \mu$, $V \in \Xi_\alpha$ we require 
\begin{equation}
0 = \int_M \mathcal{L}_U \alpha \wedge \alpha \wedge \sigma = \int (\iota_U \alpha) \left (d \alpha \wedge \sigma + d(\alpha \wedge \sigma) \right ),
\end{equation}
and since $\iota_U \alpha$ is arbitrary we find $d \alpha \wedge \sigma + d(\alpha \wedge \sigma)=0$, or
\begin{equation}
d \nu = \eta \wedge \nu.
\label{eq:condon}
\end{equation}
Any element of $\Xi_\alpha$ must then be tangent to $\mathcal{F}_\alpha$ and satisfy \eqref{eq:condon}. This is not a complete characterisation, there are vector fields satisfying these two conditions which are not elements of $\Xi_\alpha$. This is demonstrated by example in section \ref{sec:cas}. We speculate that vector fields of the form \eqref{eq:what} fully characterise $\Xi_\alpha$.

\subsection{The Poisson Bracket on $\Omega^1_I(M)$}

We define the Poisson bracket on $\Omega^1_I(M)$ which continues to take the standard form
\begin{equation}
\{F,G\}_I = \left \langle \alpha , \left [ \frac{\delta F}{\delta \alpha}, \frac{\delta G}{\delta \alpha} \right ] \right \rangle,
\label{eq:brack}
\end{equation}
where now $\alpha \in \Omega^1_I(M)$ and the functional derivatives are cosets in $\mathfrak{X}(M)/\Xi_\alpha$. The bracket must not depend on the choice of representative vector field for the functional derivative. Consider a vector field $A$ on $M$ such that $\iota_A \alpha =0$ and $d(\iota_A \mu)  = \eta \wedge \iota_A \mu $. From the previous section we know all elements of $\Xi_\alpha$ have these properties. Then
\begin{equation}
\langle \alpha , [ A,V ] \rangle = 0
\label{eq:shower}
\end{equation}
where $V \in \mathfrak{X}(M)$. We compute
\begin{equation}
\langle \alpha , [A,V ] \rangle = \int_M \alpha \wedge \iota_{[A,V]} \mu = - \int_M \alpha \wedge \mathcal{L}_V \iota_A \mu.
\end{equation}
Now $\iota_A \mu = \alpha \wedge \sigma$ for some 1-form $\sigma$. Then we have
\begin{equation}
\langle \alpha , [A,V ] \rangle =  \int_M \alpha \wedge \sigma \wedge \mathcal{L}_V \alpha = -\int_M \iota_V \alpha (2d \alpha \wedge \sigma - \alpha \wedge d \sigma ) =0,
\end{equation}
it follows that the bracket $\{F,G\}_I$ does not depend on the choice of representative vector field for the functional derivatives and becomes a Poisson bracket on $\Omega^1_I(M)$. Finally, we note that if $F$ is the restriction to $\Omega^1_I(M)$ of a functional on $\g^\ast$, then its functional derivative is still an element of $\g$, all elements of which are representative vector fields of a coset in $\mathfrak{X}(M)/\Xi_\alpha$, and the bracket \eqref{eq:brack} reproduces the Lie-Poisson bracket of the original ideal fluid formulation. In particular, we can recover Euler's equations by choosing the appropriate Hamiltonian.  

\subsection{The Godbillon-Vey Invariant}

For a codimension-1 foliation $\mathcal{F}$ on a closed manifold $M$, the Godbillon-Vey class~\cite{gv,candel} is an element $GV \in H^3(M ; \mathbb{R})$, if $M$ is a closed 3-manifold $H^3(M ; \mathbb{R}) = \mathbb{R}$ and $GV \in \mathbb{R}$ is a diffeomorphism invariant of the foliation. Let $\beta$ be a defining 1-form for $\mathcal{F}$, then the integrability condition $\beta \wedge d \beta =0$ implies there is a 1-form $\eta$ such that
\begin{equation}
d \beta = \beta \wedge \eta.
\end{equation}
The 3-form $\eta \wedge d \eta$ is closed and $GV$ is defined as
\begin{equation}
GV = \int_M \eta \wedge d \eta,
\end{equation}
$\beta$ is only defined up to multiplication by a non-zero function, and $\eta$ is only defined up to addition of a multiple of $\beta$, but under these transformations $\eta \wedge d \eta$ changes by an exact 3-form, so $GV$ is well-defined. By construction $GV$ is a diffeomorphism invariant of $\mathcal{F}$. 

\subsection{The Godbillon-Vey Invariant as a Restricted Casimir}
\label{sec:cas}
Our goal is to show that the Godbillon-Vey invariant is a Casimir with respect to the Poisson bracket $\{,\}_I$ defined above. Consider the variation of $GV$,
\begin{equation}
\frac{d  }{dt} GV \big|_{t=0}= 2 \int_M \dot{\eta} \wedge d \eta =  2 \int_M \dot{\eta} \wedge \beta \wedge \gamma.
\end{equation}
Where we use the fact that $d \eta = \alpha \wedge \gamma$. Using $d \alpha = \alpha \wedge \eta$, we find $d \dot{\alpha} = \dot{\alpha} \wedge \eta+{\alpha} \wedge \dot{\eta}$, so that
\begin{equation}
\frac{d  }{dt} GV \big|_{t=0}= 2 \int_M (\dot{\alpha} \wedge \eta - d \dot{\alpha})\wedge \gamma = \int_M \dot{\alpha} \wedge 2( \eta \wedge \gamma - d \gamma).
\end{equation}
Now we consider the 2-form $\chi = 2 (\eta \wedge \gamma - d \gamma)$, there is a freedom in $\chi$ arising from freedom in $\eta$ and $\gamma$. We may make the transformations
\begin{equation}
\eta \to \eta + f \alpha, \quad \gamma \to \gamma + f \eta - df + g \alpha,
\end{equation}
for functions $f,g$. Under these transformations one finds
\begin{equation}
\chi \to \chi - 2 ( g d \alpha + d(g \alpha)),
\end{equation}
which does not affect the value of $d GV / dt$ as per \eqref{eq:degeneracy}. Now observe that $\alpha\wedge \chi =0$ and $d\chi = \eta \wedge \chi $. Writing $\chi = \iota_T \mu$, we find $T$ is tangent to $\mathcal{F}_\alpha$ and satisfies \eqref{eq:condon}. Using \eqref{eq:shower} we find
\begin{equation}
\{F, GV \}_I =0
\end{equation}
for any functional $F$ on $\Omega^1_I(M)$ so that GV is a restricted Casimir of three-dimensional ideal fluids. 

\thanks{I am extremely grateful to PJ~Morrison for a hugely enlightening discussion. I would also like to acknowledge many useful conversations with JH~Hannay.}

\end{document}